# Power and Hydrogen Hybrid Transmission for Renewable Energy Systems: An Integrated Expansion Planning Strategy

Jin Lu, *Student Member, IEEE*, and Xingpeng Li, *Senior Member, IEEE*

*Abstract*—The increasing interest in hydrogen as a clean energy source has led to extensive research into its transmission, storage, and integration with bulk power systems. With the evolution of hydrogen technologies towards greater efficiency, and cost-effectiveness, it becomes essential to examine the operation and expansion of grids that include both electric power and hydrogen facilities. This paper introduces an expansion strategy for electric power and hydrogen transmission systems, tailored for future renewable energy-enriched grids. Our proposed transmission expansion planning with hydrogen facilities (TEP-H) model integrates daily operations of both electric power and hydrogen transmissions. The fuel cells and electrolyzers are used for electrical-hydrogen energy conversion, and related constraints are considered in TEP-H. We applied TEP-H to the Texas 123-bus backbone transmission grid (TX-123BT), for various renewable penetration levels and hydrogen technology development assumptions. It gave us insights on the scenarios that hydrogen transmission become feasible and economically beneficial. We also compared the performance of TX-123BT system with the hybrid transmission investment and the pure electrical transmission investment obtained by a traditional transmission expansion planning (TEP-T) model. The numerical results indicate that future renewable grids can have lower total cost with TEP-H if future electrical-hydrogen energy conversion efficiency is high.

## Nomenclature

*Sets*
G      Set of existing generators in the power system.
$G(p)$      Set of online generators in period $p$.
NG      Set of new generators in the power system.
H      Set of candidate hydrogen transmission routes.
T      Set of time intervals in a day.
$D^T$      Set of typical days in a year.
P      Set of future time periods studied in TEP.
NL      Set of candidate new lines.
$L^{T(n)}$      Set of transmission lines which's to bus is $n$.
$L^{F(n)}$      Set of transmission lines which's from bus is $n$.
$NL^{T(n)}$      Set of candidate new lines which's to bus is $n$.
$NL^{F(n)}$      Set of candidate new lines which's from bus is $n$.
$G^{B(n)}$      Set of generators located on bus $n$.
$NG^{B(n)}$      Set of new generators located on bus $n$.
$R^{B(n)}$      Set of renewables located on bus $n$.

*Parameters*
$B^{MVA}$      The MVA base of the grid model.
$N^P$      Number of periods in the TEP study.
$N^Y$      Number of years in each future period.
$N^D$      Number of typical days in each year.
$C_g^G$      The operation cost of generator $g$ per MWh output.
$C_g^{NG}$      The operation cost of new generator $g$ per MWh output.
$C_k^{NL}$      The construction cost of transmission line $k$.
$C_h^H$      The construction cost of hydrogen pipeline $h$.
$C_h^E$      The construction cost of electrolyzers related to hydrogen pipeline $h$.
$C_h^F$      The construction cost of fuel cells related to hydrogen pipeline $h$.
$R^M$      The ratio of the maintenance cost to the construction cost of a transmission line.
$R^{MH}$      The ratio of the maintenance cost to the construction cost of a hydrogen pipeline.
$p_g^{Min}$      The minimum output power for generator $g$.
$p_g^{Max}$      The maximum output power for generator $g$.
$p_{g,p}^{Min}$      The minimum output power limit for new generator $g$ in future period $p$.
$p_{g,p}^{Max}$      The maximum output power limit for new generator $g$ in future period $p$.
$p_r^{Min,R}$      The minimum output power for renewable $r$.
$p_{r,t,d,p}^{Max,R}$      The available power output for renewable $r$ at time interval $t$ in typical day $d$ for future period $p$.
$p_r^{Min,NR}$      The minimum output power for new renewable $r$.
$p_{r,t,d,p}^{Max,NR}$      The available power output for new renewable $r$ at time interval $t$ in typical day $d$ for future period $p$.
$p_{k,t,d,p}^{Max}$      The active power rating of line $k$ at time interval $t$ in typical day $d$ for future period $p$.
$p_{k,t,d,p}^{Max,NL}$      The active power rating of candidate new line $k$ at time interval $t$ in typical day $d$ for future period $p$.
$p_h^{E,Max}$      The power rating of electrolyzers for hydrogen transmission route $h$.
$p_h^{F,Max}$      The power rating of fuel cells for hydrogen transmission route $h$.
$h_h^{Max}$      The hydrogen transmission capacity of hydrogen pipeline for hydrogen transmission route $h$.
$p_{p,d,n,t}$      The load on bus $n$ at time interval $t$ in typical day $d$ for future period $p$.
$p_{p,d,n,t}^{SD}$      The load shedding of bus $n$ at time interval $t$ in typical day $d$ for future period $p$.
M      A large number M.
$x_k^L$      Reactance of transmission line $k$.
$x_k^{NL}$      Reactance of new transmission line $k$.
$\eta_h^E$      Efficiency of electrolyzers for hydrogen transmission route $h$.
$\eta_h^F$      Efficiency of fuel cells for hydrogen transmission route $h$.

*Variables*
$C^G$      The total operation cost of generators in future periods.

Jin Lu and Xingpeng Li are with the Department of Electrical and Computer Engineering, University of Houston, Houston, TX, 77204, USA (e-mail: jlu28@uh.edu; xingpeng.li@asu.edu).

| Symbol | Description |
|---|---|
| $C^H$ | The total investments of hydrogen facilities in future periods. |
| $C^{NL}$ | The total construction cost of transmission lines. |
| $p^G_{g,t,d,p}$ | The active power output of generator $g$ at time interval $t$ in typical day $d$ for future period $p$. |
| $p^{NG}_{g,t,d,p}$ | The active power output of new generator $g$ at time interval $t$ in typical day $d$ for future period $p$. |
| $V^{NL}_{k,p}$ | When line $k$ is constructed in future period $p$, its value is 1. Otherwise, its value is 0. |
| $V^H_{h,p}$ | When hydrogen pipeline $h$ is constructed in future period $p$, its value is 1. Otherwise, its value is 0. |
| $u^{NL}_{k,p}$ | When line $k$ is operational in future period $p$, its value is 1. Otherwise, its value is 0. |
| $u^H_{h,p}$ | When hydrogen transmission route $h$ is operational in future period $p$, its value is 1. Otherwise, its value is 0. |
| $p^L_{j,t,d,p}$ | The active power flow on line $j$ at time interval $t$ in typical day $d$ for future period $p$. |
| $p^{NL}_{j,t,d,p}$ | The active power flow on new line $j$ at time interval $t$ in typical day $d$ for future period $p$. |
| $p^R_{r,t,d,p}$ | The active power output of renewable $r$ at time interval $t$ in typical day $d$ for future period $p$. |
| $p^{NR}_{r,t,d,p}$ | The active power output of new renewables $r$ at time interval $t$ in typical day $d$ for future period $p$. |
| $p^{R,CUR}_{r,t,d,p}$ | The active power curtailment of renewable $r$ at time interval $t$ in typical day $d$ for future period $p$. |
| $p^{NR,CUR}_{r,t,d,p}$ | The active power curtailment of new renewables $r$ at time interval $t$ in typical day $d$ for future period $p$. |
| $p^E_{h,t,d,p}$ | The active power consumed by electrolyzers for hydrogen transmission route $h$ at time interval $t$ in typical day $d$ for future period $p$. |
| $p^F_{h,t,d,p}$ | The active power generated by fuel cells $f$ for hydrogen transmission route $h$ at time interval $t$ in typical day $d$ for future period $p$. |
| $p^C_{h,t,d,p}$ | The active power consumed by compressors for hydrogen transmission route $h$ at time interval $t$ in typical day $d$ for future period $p$. |
| $p^{LOAD}_{n,t,d,y}$ | The load demand on bus $n$ at time interval in typical day $d$ for future period $p$. |
| $h^H_{h,t,d,p}$ | The amount of hydrogen transmitted through hydrogen pipeline h during time interval $t$ in typical day $d$ for future period $p$. |
| $\theta^F_{k,t,d,p}$ | The phase angle of the from bus of line $k$ at time interval $t$ in typical day $d$ for future period $p$. |
| $\theta^T_{k,t,d,p}$ | The phase angle of the to bus of line $k$ at time interval $t$ in typical day $d$ for future period $p$. |

## I. Introduction

Hydrogen energy is increasingly capturing the attention of modern society, especially in the fields of transportation, energy storage, and power generation [1]-[5]. In the power industry, hydrogen is recognized as a clean energy source and is pivotal to the development of renewable grids with minimal or zero carbon emissions [6]-[7]. Hydrogen's potential as a key component in future grids is underscored by its environmentally friendly production and usage. It can be produced from water using electrolyzers, a process that consumes electric power while yielding oxygen as a byproduct—a potential resource under study for further utilization. The consumption of hydrogen, mainly through fuel cells, generates electricity for various applications like transportation and power generation. Fuel cells oxidize hydrogen, producing water and heat, which can be effectively used, underscoring hydrogen's clean and sustainable nature for future renewable energy grids.

Electric power grids can integrate hydrogen resources through a series of processes including electric-hydrogen conversion, hydrogen storage, and transmission via pipelines [8]-[10]. Electric energy in power grids is transmitted through lines from generators to loads and can be stored in various forms such as batteries, compressed air, and pumped-hydro storage. To integrate hydrogen transmission and storage into the electric power system, facilities such as fuel cells and electrolyzers are utilized to enable the conversion of hydrogen and electric energy into each other. From the electric grid's perspective, electric energy can first convert into hydrogen energy, and then be transmitted and stored through the hydrogen pipeline and storage, which will be converted back into electricity whenever it is needed. Hence, the integration of hydrogen can improve the electric grids on both transmission and storage aspects.

The future grids are expected to be dominated by renewable energy resources [11]-[12], necessitating substantial energy storage capacity to manage the intermittency and unpredictability of these sources. Moreover, such renewables-dominated power grids will require extensively additional transmission capacity, influenced by the geographical distribution of renewable energy plants [13]-[14]. Often located in areas with high wind or solar radiation potential, these plants are typically distant from load centers, necessitating the transfer of large amounts of renewable energy. Employing or integrating existing hydrogen transmissions have the potential to alleviate the high transmission capacity demands caused by significant renewable penetration. While existing hydrogen pipelines are limited, there are examples of utilities upgrading natural gas pipelines for hydrogen or hydrogen blend delivery [15]-[17]. Additionally, a number of studies are assessing the feasibility and economics of hydrogen transmission [18]-[19]. This indicates the growing potential for hydrogen transmission in real-world applications, supported by the trend towards renewable development and the existing natural gas network.

However, practical applications of hydrogen still face challenges, including high costs, leakage risks, and energy conversion inefficiencies [20]-[22]. The small size of hydrogen molecules can lead to increased leakage rates. Additionally, hydrogen embrittlement can weaken metals, necessitating pipelines made of or lined with materials resistant to this phenomenon, increasing production and maintenance costs. Moreover, energy conversion inefficiency remains a significant barrier to hydrogen integration into power systems [23]. For instance, the ideal efficiency for PEM electrolyzers is around 80% [24]-[25], and only 60% for PEM fuel cells [26]-[27]. This means a substantial portion of energy is lost during conversion, making increased efficiency crucial for hydrogen's viability in power transmission.

Despite these challenges, hydrogen transmission is expected to become more feasible and practical for future renewable energy grids. As grids incorporate more renewables, the low efficiency of conversion becomes less critical, particu-larly when utilizing abundant, low-cost renewable energy for conversion processes. Additionally, advancements in hydrogen technologies are likely to reduce costs associated with hydrogen conversion and transmission [28]-[31]. For example, the cost of electrolyzers is anticipated to decrease due to economies of scale in both module and manufacturing plant sizes [32]. The International Renewable Energy Agency (IRENA) projects that investment costs for electrolyzers could decrease by 40% in the near term and potentially by up to 80% in the future [33]. Improvements in fuel cell durability and power density are also expected. In [34], the key points to increase the power density of proton-exchange membrane fuel cell (PEMFC) is discussed. Research focused on fuel cell catalyst technology is pivotal for enhancing fuel cell durability [35]-[36]. For hydrogen transmission, materials that slow down hydrogen embrittlement and reduce leakage risks are under development, with numerous studies investigating suitable materials for hydrogen pipelines [37]-[39].

While significant advancements in hydrogen technologies are anticipated, the operation and planning of renewable grids integrated with hydrogen remain critical for optimizing the advantages and benefits of hydrogen. Current strategies employed by independent system operators (ISOs) or regional transmission organizations (RTOs), such as economic dispatch and security-constrained unit commitment (SCUC), primarily address electric transmission constraints like nodal power balance and line flow equations [40]-[41]. However, these strategies do not account for hydrogen transmission constraints, rendering them insufficient for managing hydrogen facility operations. Furthermore, there is a lack of consideration for the synergy between electric and hydrogen transmission systems, which could yield optimal energy transmission solutions for grid benefits. It's also necessary for grid planning to consider hydrogen integration. Wise investment in hydrogen facilities is essential to achieve the reliability and economic efficiency of the power-hydrogen hybrid energy grids. Suitable facility placement lays the foundation for reliable and economical grid operations. Expansion planning strategies should account for operational conditions across different future periods, encompassing constraints similar to those in operational strategies, including generation capacity ramping, reserve requirements, and transmission constraints like line flow thermal limits. Unlike the day-ahead or real-time operational strategy, transmission planning involves considering multi-scenario forecasts, accommodating long-term shifts in generation and load. This planning approach may involve investing in new transmission lines in one area and hydrogen pipelines in another, depending on future conditions, to optimize grid benefits.

The integration of hydrogen into power grids is still nascent, with pioneering research exploring its feasibility and economics [42]-[43]. Studies on operational strategies for hydrogen-integrated power systems have begun to emerge. For instance, [45] models and simulates the daily operation of a renewable grid with hydrogen transmission or hydrogen hubs, while [46] proposes a seasonal operation strategy for a renewable grid with salt caverns. However, literature on the planning of renewable grids with hydrogen transmission remains limited. To address this gap, we propose the Transmission Expansion Planning for Hydrogen-Integrated Grids (TEP-H). This strategy aims to identify optimal investments that are cost-effective and adhere to the physical constraints of renewable grids and hydrogen facilities, including pipelines, fuel cells, electrolyzers, and compressors. TEP-H takes into account capital and maintenance costs of electric transmission lines, thermal generation costs, and the capital and maintenance costs of hydrogen pipelines, fuel cells, and electrolyzers. We apply TEP-H to the synthetic Texas 123-bus backbone transmission grid (TX-123BT), which features detailed future renewable and load profiles, and consider various improvements in hydrogen technology in our simulations. The numerical results are thoroughly analyzed, and the main contributions of this work are summarized as follows:

- This study is the very first to propose the planning model for renewable energy grids with both electric and hydrogen transmissions.
- The proposed TEP-H models are explored under various scenarios of hydrogen technology improvements, determining when and under what circumstances hydrogen transmission becomes feasible and cost-effective.
- The operation conditions of future hybrid energy grids with integrated power-hydrogen transmissions are analyzed and contrasted with those of a purely electric transmission grid.

The remainder of this work is organized as follows: Section II elucidates hydrogen transmission and its cooperation with electric transmission. Section III details traditional TEP and the proposed TEP-H formulations. Section IV examines the future operation conditions under hybrid and pure electric transmission investments, and also explores the impact of conversion efficiency improvement and cost reduction on the participation of hydrogen transmission in future grids. Conclusions are drawn in Section V.

## II. COOPERATION OF HYDROGEN AND ELECTRIC TRANSMISSION IN RENEWABLE GRIDS

In power grids with high renewable energy penetration, substantial transmission capacity is required to carry renewable energy to load areas. Anticipating a significant increase in renewable power plants in the near future, the transmission capability of current power grids needs substantial enhancement. In the electrical transmission network, renewable energy is transferred through transmission lines. Conversely, in the hydrogen transmission network, electrical energy is first converted into hydrogen energy via electrolyzers, then transmitted through hydrogen pipelines. Near the load areas, the hydrogen energy is reconverted into electrical energy through fuel cells. Fig. 1 illustrates the pathways of electric, hydrogen, and hybrid transmission configurations.

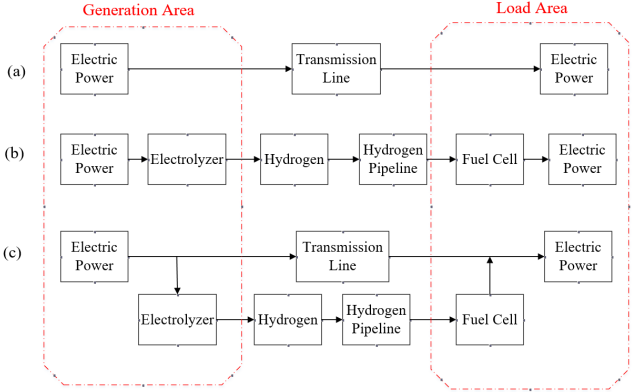

Fig. 1. Different Transmission Configurations: (a) Electrical Transmission (b) Hydrogen Transmission (c) Hybrid Transmission.

Electrical transmission energy losses are mainly Joule losses, which are 2-3% of the total power transmitted [47]. In contrast, hydrogen transmission, involving electric-to-hydrogen energy conversion, experiences round-trip conversion losses exceeding 50% under ideal conditions. Therefore, in current scenarios, electrical transmission is often more practical and advantageous than hydrogen transmission. However, in future grids with excess renewable energy that may be simply curtailed and thus wasted, the significance of transmission losses may diminish, making hydrogen or hybrid transmission viable alternatives. These methods can ensure utilization of renewable energy that might otherwise be curtailed. Hybrid transmission leverages both electric and hydrogen transmission facilities, potentially achieving a higher maximum transfer capacity — the sum of the capacities of both electrical and hydrogen transmission. This configuration becomes increasingly relevant when existing transmission lines require additional capacity, which can be supplemented by new hydrogen transmission facilities during expansion.

Future renewable grids may employ these three transmission configurations to expand transmission capabilities in various local areas and scenarios. Electric transmission investments might be prioritized where significant capacity is needed, and less loss of transmission is required. Hydrogen transmission could be more suitable in areas where existing natural gas pipelines can be upgraded for hydrogen transmission or for long-distance transmission from renewable generation sites, depending on advancements in hydrogen technology. Hybrid transmission could be implemented when a new hydrogen pipeline is constructed alongside an existing electrical transmission line. Moreover, determining optimal sending and receiving locations for transmission is not straightforward. Reference [45] suggests that the ideal receiving location for hydrogen transmission, also the location of fuel cells, may be a transitional area between load and generation zones. An example follows to elucidate the concept of transmission expansion with hybrid transmission and to illustrate the scenarios where such transmission is beneficial.

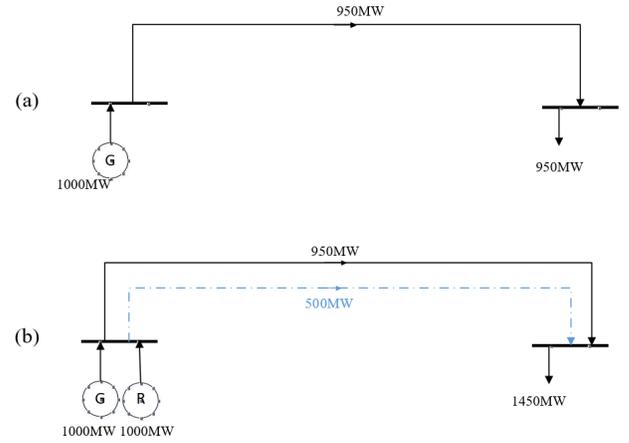

Fig. 2. (a) An example of point-to-point transmission (b) The future scenario with hydrogen transmission investment.

In Fig. 2. (a), an electrical transmission line transfers the electric power to meet 950 MW load demand. In the future, the load demand is expected to reach 1,450 MW in some peak load conditions, and 1,000 MW renewables are deployed in the generation area. In this case, we can invest in the transmission facilities illustrated in Fig. 1. (b), which is equivalent to transferring 500 MW electrical power and it is represented by the blue line in Fig. 2. (b). In this case, we assume the round-trip efficiency of hydrogen transmission is 50%, and the increased load is covered by the developed renewables and the hydrogen transmission. With this transmission investment, the combined transmission in Fig. 2 (b) is considered a form of hybrid transmission, as depicted in Fig. 1 (c). If there is an existing natural gas line between the two buses, this transmission investment becomes even more economically advantageous.

During periods of low demand, for certain seasons or times of day, hydrogen facilities can be used to produce hydrogen, utilizing surplus renewable energy. This hydrogen can then supply both bus locations. Fig. 3 demonstrates this scenario, where the natural gas power plant operates at its minimum output, and renewable energy covers additional loads not met by natural gas generation, while also producing hydrogen for both locations.

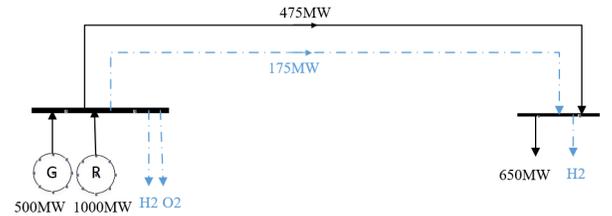

Fig. 3. In low demand time, the hydrogen facilities can provide H2 to both locations.

## III. Transmission Expansion Planning with Hydrogen Transmission Facilities

Transmission expansion is a crucial long-term planning problem aimed at identifying optimal transmission investment for future periods. The future grids upgraded with the trans-mission investments should remain reliable throughout all planned periods, and minimize instances of load shedding. Reliability, while fundamental to TEP, is not the sole consideration. Cost-effectiveness is equally important, encompassing both the system-wide operational costs and the capital costs of new transmission investments. In TEP-H, the planning encompasses multiple representative periods, each spanning $N^Y$ years, under the assumption that grid conditions remain relatively consistent throughout these years. Each period includes four typical days, symbolizing different seasonal conditions within a year. The objective function of TEP-H, as shown in (1), comprises various cost components: the production cost of thermal generator $C^G$, the capital cost of new transmission line $C^{NL}$, the capital cost of hydrogen facilities $C^H$, and a penalty term for the load sheddings. The penalty term includes the load shedding amount along all the periods $p^{SD}_{b,t,d,p}$, and a big number M as weight coefficient.

The thermal generator production cost $C^G$ accounts for the total operation costs of both existing and future generators, as detailed in (2). The total cost of new lines $C^{NL}$ outlined in (3), includes both initial capital and ongoing maintenance costs. Here, $V^{NL}_{k,p}$ is a binary variable indicating the construction of a candidate line $k$ during a specific period $p$. Once constructed, line $k$ incurs maintenance costs for $N^P - p + 1$ periods, with these costs being proportional to the initial capital expense. (4) details the capital costs associated with hydrogen facilities, encompassing pipelines, fuel cells, and electrolyzers, applicable when a candidate hydrogen pipeline is slated for construction.

$$\min C^{NL} + C^G + C^H + M * \sum_{n \in N, t \in T, d \in D^T, p \in P} p^{SD}_{b,t,d,p} \quad (1)$$

$$C^G = N^Y * B^{MVA} * \frac{365}{N^D} * \sum_{g \in G, t \in T, d \in D^T, p \in P} p^G_{g,t,d,p} * C^G_g$$
$$+ N^Y * B^{MVA} * \frac{365}{N^D} * \sum_{g \in NG, t \in T, d \in D^T, p \in P} p^{NG}_{g,t,d,p} * C^{NG}_g \quad (2)$$

$$C^{NL} = \sum_{k \in NL, p \in P} V^{NL}_{k,p} * C^{NL}_k * (1 + (N^P - p + 1) * R^M * N^Y) \quad (3)$$

$$C^H = \sum_{h \in H, p \in P} V^H_{h,p} * C^H_h * (1 + (N^P - p + 1) * R^{MH} * N^Y) + \sum_{h \in H, p \in P} V^H_{h,p} * C^F_h + \sum_{h \in H, p \in P} V^H_{h,p} * C^E_h \quad (4)$$

In an electrical transmission system, each node should be power balanced. This balance is expressed by (5). On each node, the power $p^L_{j,t,d,p}$ flows along transmission lines connected to it. Additionally, power flowing on new transmission lines, represented by $p^{NL}_{j,t,d,p}$, is also taken into account. The power generated by existing and new generators or renewable power plants are symbolized by $p^G_{g,t,d,p}$, $p^{NG}_{g,t,d,p}$, $p^R_{r,t,d,p}$, $p^{NR}_{n,t,d,p}$. Given the possibility of high renewable energy penetration in future grids, there may be instances where renewable production has to be curtailed to ensure matching between system load and generation. This curtailment is represented by $p^{RCUR}_{r,t,d,p}$ and $p^{NR,CUR}_{n,t,d,p}$. The power generated or consumed by fuel cell, electrolyzers, and compressors are represented by $p^F_{h,t,d,p}$, $p^E_{h,t,d,p}$, and $p^C_{h,t,d,p}$. The load demand and shedding are represented by $p^{LOAD}_{n,t,d,p}$ and $p^{SD}_{n,t,d,p}$.

$$\sum_{j \in L^T(n)} p^L_{j,t,d,p} - \sum_{j \in L^F(n)} p^L_{j,t,d,p} + \sum_{j \in NL^T(n)} p^{NL}_{j,t,d,p}$$
$$- \sum_{j \in NL^F(n)} p^{NL}_{j,t,d,p} + \sum_{g \in G^B(n)} p^G_{g,t,d,p} + \sum_{g \in NG^B(n)} p^{NG}_{g,t,d,p}$$
$$+ \sum_{r \in R^B(n)} p^R_{r,t,d,p} + p^{NR}_{n,t,d,p} - \sum_{r \in R^B(n)} p^{RCUR}_{r,t,d,p} - p^{NR,CUR}_{n,t,d,p} \quad (5)$$
$$+ \sum_{h \in H^F(n)} p^F_{h,t,d,p} - \sum_{h \in H^E(n)} p^E_{h,t,d,p} - \sum_{h \in H^C(n)} p^C_{h,t,d,p}$$
$$= p^{LOAD}_{n,t,d,p} - p^{SD}_{n,t,d,p}$$
$$\forall n \in B, t \in T, d \in D^T, p \in P$$

The maximum amount of load shedding and the renewable curtailment are described by (6)-(8). Besides, the maximum thermal generation output power and the renewable production are shown in (9)-(12).

$$0 \leq p^{SD}_{b,t,d,p} \leq p^{LOAD}_{n,t,d,p} \quad \forall n,t,d,p \in N,T,D^T,P \quad (6)$$

$$0 \leq p^{RCUR}_{r,t,d,p} \leq p^R_{r,t,d,p} \quad \forall r,t,d,p \in R,T,D^T,P \quad (7)$$

$$0 \leq p^{NR,CUR}_{r,t,d,p} \leq p^{NR}_{r,t,d,p} \quad \forall r,t,d,p \in NR,T,D^T,P \quad (8)$$

$$p^{Min}_g \leq p^G_{g,t,d,p} \leq p^{Max}_g \quad \forall g,t,d,p \in G,T,D^T,P \quad (9)$$

$$p^{Min}_{g,p} \leq p^{NG}_{g,t,d,y} \leq p^{Max}_{g,p} \quad \forall g,t,d,p \in NG,T,D^T,P \quad (10)$$

$$p^{Min,R}_r \leq p^R_{r,t,d,p} \leq p^{Max,R}_{r,t,d,p} \quad \forall r,t,d,p \in R,T,D^T,P \quad (11)$$

$$p^{Min,NR}_r \leq p^{NR}_{r,t,d,y} \leq p^{Max,NR}_{r,t,d,p}$$
$$\forall r \in NR, t \in T, d \in D^T, p \in P \quad (12)$$

$$-p^{Max}_{k,t,d,p} \leq p^L_{k,t,d,p} \leq p^{Max}_{k,t,d,p}$$
$$\forall k \in L, t \in T, d \in D^T, p \in P \quad (13)$$

$$p^L_{k,t,d,p} = \frac{\theta^F_{k,t,d,p} - \theta^T_{k,t,d,p}}{x^L_k}$$
$$\forall k \in L, t \in T, d \in D^T, p \in P \quad (14)$$

$$-M * (1 - u^{NL}_{k,p}) \leq p^{NL}_{k,t,d,p} - \frac{\theta^F_{k,t,d,p} - \theta^T_{k,t,d,p}}{x^{NL}_k} \leq M *$$
$$(1 - u^{NL}_{k,p}), \forall k \in NL, t \in T, d \in D^T, p \in P \quad (15)$$

$$-p^{Max,NL}_{k,t,d,p} * u^{NL}_{k,p} \leq p^{NL}_{k,t,d,p} \leq p^{Max,NL}_{k,t,d,p} * u^{NL}_{k,p}$$
$$\forall k \in NL, t \in T, d \in D^T, p \in P \quad (16)$$

$$\sum_{p' \in P, p' \leq p} u^{NL}_{k,p'} \leq u^{NL}_{k,p} \quad \forall k \in NL, p \in P \quad (17)$$

$$v^{NL}_{k,p} \geq u^{NL}_{k,p} - u^{NL}_{k,p-1}, \forall k \in NL, p \in P, p > 1 \quad (18)$$

$$v^{NL}_{k,1} = u^{NL}_{k,1} \quad \forall k \in NL \quad (19)$$

$$\eta^E_h * p^E_{h,t,d,p} * B^{MVA} = h^H_{h,t,d,p} \quad (20)$$

$$\forall h \in H, t \in T, d \in D^T, p \in P$$

$$\eta_h^F * h_{h,t,d,p}^H * B^{MVA} = p_{h,t,d,p}^E \quad (21)$$
$$\forall h \in H, t \in T, d \in D^T, p \in P$$

$$\eta_h^C * h_{h,t,d,p}^H = p_{h,t,d,p}^C * B^{MVA} \quad (22)$$
$$\forall h \in H, t \in T, d \in D^T, p \in P$$

$$h_{h,t,d,p}^H \leq h_h^{Max} * u_{h,y}^H \quad (23)$$
$$\forall h \in H, t \in T, d \in D^T, p \in P$$

$$p_{h,t,d,p}^E \leq p_h^{E,Max} * u_{h,y}^H \quad (24)$$
$$\forall h \in H, t \in T, d \in D^T, p \in P$$

$$p_{h,t,d,p}^F \leq p_h^{F,Max} * u_{h,y}^H \quad (25)$$
$$\forall h \in H, t \in T, d \in D^T, p \in P$$

$$\sum_{p' \leq p} v_{h,p'}^H \leq u_{h,p}^H \quad \forall h \in H \quad (26)$$

$$v_{h,p}^H \geq u_{h,p}^H - u_{h,p-1}^H, \forall h \in H, p \in P, p > 1 \quad (27)$$

$$v_{h,1}^H = u_{h,1}^H \quad \forall h \in H \quad (28)$$

Power flow on transmission lines must adhere to thermal limits as outlined in (13)-(14). For candidate transmission lines, power flow constraints become applicable only after their construction, as specified in (15). Equation (16) ensures that the maximum power flow on an unbuilt candidate line is zero. The binary variables governing candidate transmission lines are further constrained by (17)-(19). Efficiency of conversions between electrical and hydrogen energy, involving electrolyzers and fuel cells, is modeled through (20)-(21). Moreover, compressors consume electrical energy to pressurize hydrogen for pipeline transmission, as detailed in (22). Hydrogen pipelines, electrolyzers, and fuel cells have maximum transmission and working power capacities, respectively, as described in (23)-(25). Constraints for binary variables associated with candidate hydrogen facilities are laid out in (26)-(28).

The TEP-H model takes into account long-term variations in renewable production $p_{r,t,d,p}^{Max,R}$, load $p_{n,t,d,p}^{LOAD}$, and dynamic line ratings $p_{k,t,d,p}^{Max}$ over the planning period. Based on these projections, TEP-H determines the construction and online status of both electric and hydrogen transmission investments, represented by variables $v_{k,p}^{NL}$, $u_{k,p}^{NL}$, $v_{h,p}^H$, and $u_{h,p}^H$. The model outputs future grid operation conditions, including renewable curtailment $p_{r,t,d,p}^{RCUR}$, load shedding $p_{b,t,d,p}^{SD}$, generation dispatching $p_{g,t,d,p}^G$, and both electric power flow $p_{k,t,d,p}^L$ and hydrogen flow $h_{h,t,d,p}^H$ in the hybrid system for various scenarios. It also calculates the system's operational and investment costs, including thermal generator production cost $C^G$, new transmission line capital cost $C^{NL}$, and hydrogen facilities capital cost $C^H$, thereby providing a comprehensive view of transmission investments, future grid operations, and associated costs.

To validate TEP-H and compare transmission investments with and without hydrogen facilities, a traditional TEP model (TEP-T) is also formulated. TEP-T, similar in structure to TEP-H but excluding hydrogen-related constraints, omits hydrogen facility costs in its objective function (29) and excludes electrical energy generated or consumed by fuel cells and electrolyzers in its power balance equation (30). The complete formulations of both TEP-T and TEP-H are summarized in Table I.

$$\min C^{NL} + C^G + M * \sum_{n \in N, t \in T, d \in D^T, p \in P} p_{b,t,d,p}^{SD} \quad (29)$$

$$\sum_{j \in L^{T(n)}} p_{j,t,d,p}^L - \sum_{j \in L^{F(n)}} p_{j,t,d,p}^L + \sum_{j \in NL^{T(n)}} p_{j,t,d,p}^{NL}$$
$$- \sum_{j \in NL^{F(n)}} p_{j,t,d,p}^{NL} + \sum_{g \in G^{B(n)}} p_{g,t,d,p}^G + \sum_{g \in NG^{B(n)}} p_{g,t,d,p}^{NG}$$
$$+ \sum_{r \in R^{B(n)}} p_{r,t,d,p}^R + p_{n,t,d,p}^{NR} - \sum_{r \in R^{B(n)}} p_{r,t,d,p}^{RCUR} - p_{n,t,d,p}^{NR,CUR} \quad (30)$$
$$= p_{n,t,d,p}^{LOAD} - p_{n,t,d,p}^{SD}$$
$$\forall n \in B, t \in T, d \in D^T, p \in P$$

TABLE I
THE MODEL FORMULATIONS OF TEP-T AND TEP-H.

| Proposed TEP-H | Benchmark TEP-T |
| --- | --- |
| (1)-(28) | (2)-(3), (6)-(19), (29)-(30) |

IV. CASE STUDIES

The effectiveness of TEP-H, like many other planning strategies, hinges on reliable prediction data for different grid sections. However, due to the sensitivity and confidentiality of power system data, obtaining such information can be challenging or necessitate confidential authorization. A notable exception is the publicly available dataset such as the TX-123BT test system, as explained in [48]. This dataset provides detailed renewable and thermal generation data at the facility level, along with transmission network and dynamic line rating data at the nodal or line level. Consequently, TX-123BT serves as the test system for our TEP-H simulations. We conducted the TEP-H simulations using Python with the Pyomo package. As TEP-H is a mixed-integer linear programming (MILP) problem, we employed the Gurobi solver, setting the optimality gap at 0.1%.

The future grid is fraught with uncertainties that may significantly impact its expansion. The TX-123BT's future profile is based on the Representative Concentration Pathway (RCP) 8.5 scenario [49], reflecting a high carbon emission future if current social policies remain unchanged. This scenario informs our projections of future solar and wind power plant production and dynamic line ratings. Additionally, TEP-H necessitates predictions of future hydrogen-related conditions, such as capital costs for hydrogen facilities and the efficiency of fuel cells and electrolyzers. These parameters critically influence the investment decisions of hydrogen transmission and the co-expansion of hydrogen and electric transmission system. Hydrogen technology, being state-of-the-art and rapidly evolving, presents challenges in accurately predicting future developments and conditions. Therefore, we conducted TEP-H simulations under various hydrogen technology development scenarios, providing insights into how advancements in hydrogen technology might influence the expansion of future renewable grids. A key aspect of hydrogen improvements is the round-trip efficiency, representing the efficiency of converting electrical energy to hydrogen and then back to electrical energy via electrolyzers and fuel cells. For instance, with an electrolyzer efficiency of 60% and a fuel cell efficiency of 80%, the round-trip efficiency would be 48%. Since renewable energy may offset the low efficiency issues of the hydrogen transmission, renewable penetration is also an important factor for TEP with hydrogen transmission investment.

Our first investigation focused on identifying the renewable penetration level at which hydrogen becomes a feasible solution for future grid transmission. We ran TEP-H simulations on the TX-123BT with varying levels of renewable penetration, assuming a round-trip efficiency of 40%, indicative of the most ideal current hydrogen technology. The findings, presented in Table II, suggest that hydrogen transmission becomes more viable in grids with high renewable penetration. Under the most ideal conditions for current hydrogen technology, hydrogen transmission is feasible at a renewable penetration level of around 60%.

TABLE II
THE HYDROGEN TRANSMISSION INVESTMENT FOR TX-123BT WITH VARIOUS RENEWABLE PENETRATIONS

| Renewable Penetration Level | Number of Hydrogen Pipelines | Construction Period | Total Hydrogen Investment (Million $) |
|---|---|---|---|
| 40% | 0 | N/A | 0 |
| 60% | 1 | 2036-2040 | 2,050 |
| 80% | 1 | 2031-2035 | 2,065 |

In our second set of simulations, we explored the TEP-H model under various assumptions regarding advancements in hydrogen technology, focusing on aspects such as round-trip efficiency and the costs of hydrogen facilities. Notably, we discovered that the efficiency of fuel cells and electrolyzers plays a more critical role in the application of hydrogen transmission compared to reductions in capital and maintenance costs of hydrogen facilities. This is because conversion efficiency is directly tied to the potential waste of a large amount of generated energy, which poses a greater economic burden than the costs associated with hydrogen facilities. Our findings, detailed in Tables III-IV, illustrate the investment trends in hydrogen transmission against different round-trip efficiency levels. Notably, higher round-trip efficiency leads to increased investment in hydrogen transmission, particularly when efficiency levels approach 70%-80%.

TABLE III
THE HYDROGEN INVESTMENTS FOR VARIOUS ROUND-TRIP EFFICIENCY WHEN RENEWABLE PENETRATION LEVEL IS 80%

| Round Trip Efficiency | Number of Hydrogen Pipelines | Construction Period | Total Hydrogen Investment (Million $) |
|---|---|---|---|
| 40% | 1 | 2031-2035 | 2,065 |
| 50% | 1 | 2031-2035 | 2,065 |
| 60% | 1 | 2031-2035 | 2,065 |
| 70% | 2 | 2031-2035, 2036-2040 | 4,127 |
| 80% | 4 | 2021-2025, 2031-2035, 2036-2040, 2031-2035 | 8,418 |

TABLE IV
THE HYDROGEN INVESTMENTS FOR VARIOUS ROUND-TRIP EFFICIENCY WHEN RENEWABLE PENETRATION LEVEL IS 60%

| Round Trip Efficiency | Number of Hydrogen Pipelines | Construction Period | Total Hydrogen Investment (Million $) |
|---|---|---|---|
| 40% | 1 | 2036-2040 | 2,050 |
| 50% | 1 | 2041-2045 | 2,035 |
| 60% | 1 | 2031-2035 | 2,065 |
| 70% | 1 | 2036-2040 | 2,050 |
| 80% | 3 | 2036-2040, 2041-2045, 2036-2040 | 6,201 |

The TEP-H simulations under various round-trip efficiency and cost assumptions revealed that when round-trip efficiency is low, the cost of hydrogen facilities has minimal impact on hydrogen transmission investment decisions. However, as the round-trip efficiency escalates to 70%-80%, cost reduction becomes a significant factor influencing hydrogen investment decisions. The investments under various cost reduction scenarios are presented in Table V.

TABLE V
THE HYDROGEN INVESTMENTS FOR VARIOUS COST REDUCTION WITH 60% RENEWABLE PENETRATION LEVEL AND 70% ROUND-TRIP EFFICIENCY

| Hydrogen Facilities Cost Reduction | Number of Hydrogen Pipelines | Total Hydrogen Investment (Million $) |
|---|---|---|
| 0% | 2 | 4,127 |
| 20% | 2 | 3,301 |
| 40% | 2 | 2,476 |
| 60% | 3 | 2,514 |
| 80% | 3 | 1,257 |

We also compared the future grid conditions under both TEP-T and TEP-H investments, focusing on scenarios where the renewable penetration level is high (60%). For this comparison, we assumed a 40% round-trip efficiency in the TEP-H simulation. The results of this simulation, as illustrated in Table VI, show that TEP-H involves investment in hydrogen transmission and requires less electrical transmission investment compared to TEP-T. The total cost for TEP-H amounted to 473,965 million dollars, which is 3,029 million dollars less than that of TEP-T.

TABLE VI
THE TRANSMISSION INVESTMENTS BY TEP-H AND TEP-T

| TEP Model | TEP-T | TEP-H |
|---|---|---|
| Number of Hydrogen Pipeline | N/A | 1 |
| Hydrogen Investments (M$) | N/A | 2,050 |
| Transmission Line Number | 11 | 8 |
| Transmission Investments (M$) | 4,285 | 3,433 |
| Generation Costs (M$) | 469,680 | 465,453 |
| Total Costs (M$) | 473,965 | 470,936 |

Further analysis was conducted on the grid operation conditions of the TX-123BT under TEP-H investments, with a scenario where renewable penetration is very high (80%) and hydrogen technology has significantly improved (80% round-trip efficiency). The detailed expansion planning for hydrogen transmission as determined by TEP-H is displayed in Table VII. The various associated costs are listed in Table VIII. Notably, the generation cost, accounting for planning over 30 years and all thermal power plants in the ERCOT, is substantially larger than the transmission investment cost.

TABLE VII
THE PLANNING OF HYDROGEN TRANSMISSION

| Hydrogen Pipeline Number | Construction Period | From Bus Number | To Bus Number |
|---|---|---|---|
| 1 | 2021-2025 | 98 | 1 |
| 2 | 2031-2035 | 98 | 1 |
| 3 | 2036-2040 | 91 | 31 |
| 6 | 2031-2035 | 78 | 63 |

TABLE VIII
THE OPERATION AND INVESTMENT COSTS

| Total Cost | Generation Cost | Hydrogen Facilities Investment |
|---|---|---|
| 467.43B | 459.02B | 8.41B |

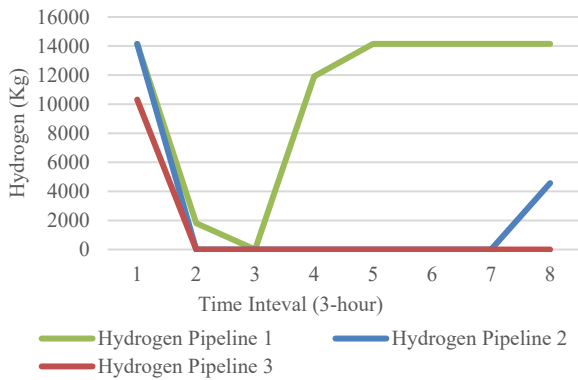

Fig. 4. Daily Operation of Hydrogen Pipelines in Quarter I, 2046-2050.

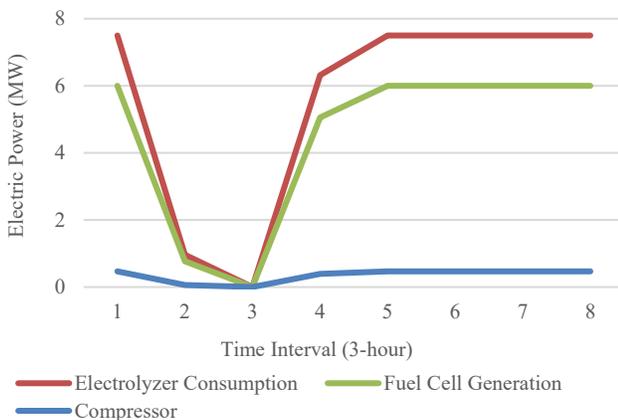

Fig. 5. Daily Operation of Fuel Cells, Electrolyzers and Compressors in Quarter I, 2046-2050.

TEP-H also provides scheduled operations for hydrogen transmission. The daily hydrogen transmission and associated operations of fuel cells and electrolyzers for a typical day in Quarter I, spanning the years 2046-2050, are depicted in Fig. 4 and Fig. 5.

## V. CONCLUSION

This paper delves into the expansion planning for renewable energy grids, with a particular focus on integrating both electric and hydrogen transmissions. The proposed TEP-H model accounts for hydrogen facilities related physical constraints and long-term shifts in future grids, enabling the determination of an optimal investment plan for both electric and hydrogen transmission. This approach aims to minimize the total system costs, encompassing both operational expenses and expansion investments, while striving to maintain system reliability. The proposed TEP-H model incorporates constraints of both power grids and hydrogen facilities, including hydrogen pipelines, electrolyzers, and fuel cells. By including capital and maintenance costs of these hydrogen facilities, TEP-H can evaluate the hydrogen transmission from the economic aspect.

Since the hydrogen related costs and the conversion efficiency are modeled in TEP-H, we are able to figure out the how the hydrogen technology improvement will influence the future hydrogen integration in the renewable grids. Based on the numerical results of TEP-H simulations on the TX-123BT, we confirm that electric-hydrogen energy conversion efficiency is the key point for the hydrogen to be applied to the future grids. TEP-H lays the groundwork and can serve as a benchmark for other planning strategies in hydrogen integration into renewable grids. It also paves the way for the industry level prospecting on practical applications of hydrogen in renewable enriched future grids.